\documentclass[iop,apj,twocolumn]{aastex7}
\usepackage{amsmath,amssymb,amstext}
\usepackage{epstopdf}

\usepackage{xcolor}

\newcommand{\msun}{M$_\odot$}

\shorttitle{Explosive outflows}
\shortauthors{Rivera-Ortiz, P. R. et al.}
\usepackage{hyperref}
\begin{document}

\title{Gravitationally Induced Explosive Outflows}

\author[0000-0001-6867-4210]{P. R. Rivera-Ortiz}
\affiliation{
Instituto de Radioastronomía y Astrofísica, Universidad Nacional Aut\'onoma de M\'exico, 58090, Morelia, Michoac\'an, M\'exico
}
\email[show]{p.rivera@irya.unam.mx}

\author{A. Rodríguez-González}
\affiliation{Instituto de Ciencias Nucleares, 
Universidad Nacional Aut\'onoma de M\'exico, 
Ap. 70-543, 04510 D.F., M\'exico}
\email{p.rivera@irya.unam.mx}

\author{J. Cantó}
\affiliation{Instituto de Astronom\'{\i}a, Universidad
Nacional Aut\'onoma de M\'exico, Ap. 70-264, 04510 D.F., M\'exico} 
\email{p.rivera@irya.unam.mx}

\author{L. A. Zapata}
\affiliation{
Instituto de Radioastronomía y Astrofísica, Universidad Nacional Aut\'onoma de M\'exico, 58090, Morelia, Michoac\'an, M\'exico
}
\email{p.rivera@irya.unam.mx}

\author{L. Hernández-Martínez}
\affiliation{ Facultad de Ciencias.  Universidad Nacional Aut\'onoma de M\'exico, Ap. 70-399, 04510 D.F., M\'exico.}
\email{p.rivera@irya.unam.mx}

\author{E. Guzmán Ccolque}
\affiliation{
 Instituto Argentino de Radioastronomía (CCT-La Plata, CONICET, UNLP, CICPBA) C.C. No. 5, 1894, Villa Elisa, Buenos Aires, Argentina
}
\email{p.rivera@irya.unam.mx}

\author{M. Fernández-López}
\affiliation{
 Instituto Argentino de Radioastronomía (CCT-La Plata, CONICET, UNLP, CICPBA) C.C. No. 5, 1894, Villa Elisa, Buenos Aires, Argentina
}
\email{p.rivera@irya.unam.mx}

% \author[0000-0002-0786-7307]{Pedro Ruben Rivera-Ortiz},
\begin{abstract}
{Over the past decade, there has been a significant increase in the reporting of extensive and luminous star-forming regions associated with explosive outflows. Nevertheless, there is still a lack of understanding of the possible physical mechanisms that produce such energetic and isotropic events. Then, we propose a gravitational interaction as a likely mechanism that could trigger explosive outflows in dense {star-forming} regions. This could constrain the physical conditions that generate an explosive outflow produced by the close dynamical encounter of a runaway star with a clump cluster in dynamical equilibrium.}
   {Then, we have produced a set of $N$-body simulations that account for the collision of a 10M$_\odot$ stellar object with a cluster of particles with a mass that ranges   from 0 to 50 \msun. We propose a parameter to describe the interaction, the evaporation parameter, that represents the fraction of stars that become unbound.}
   {The main result is that, when the cluster mass {  is less than, or up to a few times the stellar mass}, the collision will produce an explosive outflow, ejecting a significant fraction of the cluster members with velocities larger than the impact velocity.}
   {All of our models produce an explosive outflow, with different characteristics, which increases the probability that a close encounter could be responsible for producing the observed flows. }
\end{abstract}

\section{Introduction}
\label{sec:Intro}

Young star clusters are the nurseries of most stars and play an essential role in our understanding of the star formation process, as they provide hints of the parent molecular cloud mass distribution \citep{Lada03,Mckee07,Longmore14}. During the fragmentation and collapse of such molecular clouds, it is expected that the densest regions will suffer close encounters with stellar objects that can lead to the ejection of a fraction of them as runaway objects. Evidence has shown that explosive outflows, such as the one in the {Orion Becklin-Neugebauer/ Kleinman-Low \citep[Orion BN/KL,][]{Kleinmann67,Becklin67,AllenBurton1993}} region, are a consequence of this kind of interaction \citep{Zapata2009,Zapata2023}.

Explosive outflows (hereafter EO) found in 
 star-forming regions have brought a new level of complexity to molecular outflows generated by forming stars that have a bipolar structure and are generated by jets and winds. 
These EOs are observed in massive star-forming regions, challenging our understanding of the star formation process of massive stars, {and they are composed of a large number of dense clumps leading gaseous filaments that spread almost isotropically around a common center} 
\citep{Bally2017,Zapata2017,Zapata2023}. Additionally, over the last decade, thanks to the higher resolution and sensitivity of the ALMA, the number of EOs found has increased to six: Orion~BN/KL \citep{Bally2020}, DR21 \citep{Zapata2013}, G5.89-0.39 \citep{Zapata2020}, IRAS~16076-5134 \citep{Ccolque2022}, Sh106-IR and IRAS~12326-6245 \citep{Zapata2023}. Sh106-IR has recently been proposed as another EO \citep{Bally2022}, but  associated CO emissions have not been reported yet. For the case of DR21, there is still the notion that this flow might be in some way a classical bipolar outflow \citep{Skretas2023}, however its explosive nature was proven very recently \citep[][]{Guzman2024}. These outflows are characterized by a large number of high-velocity, collimated streamers or filaments driven from a common center. They are radially expanding and resemble the wakes produced by the shrapnel of a dispersive explosion. In addition, they exhibit a behavior similar to a Hubble law in which the velocity detected along a filament increases linearly with the distance from the origin of the explosion, reaching speeds of several hundred kilometers per second and their kinetic energies are of the order of $10^{48}$~erg \citep{Zapata2017}. The current number of discovered EOs and their assumed duration of $\sim1,500$~yr has set this rate to a value similar to the Galactic supernovae rate and the Galactic rate of formation of massive stars \citep{Ccolque2022, Zapata2023}. This points EOs as a more usual phenomenon, which may be regarded as an evolutionary stage of at least half of the Galactic massive stars, that suggests this phenomenon may be more common and potentially influential than previously thought.

Explosive outflows likely involve $N$-body interactions, either leading to the formation of compact binaries or stellar mergers, plus some ejected stars. In this model, their energy comes from the gravitational potential released by the binary or protostellar merger formation. However, it has not yet been possible to determine the origin of these explosive outflows, as the evidence has not been conclusive for any of the models, although there are a few proposals in the literature \citep{Bally2011,Raga2021,Rivera2021,RodriguezGonzalez23}. {  It is uncertain if only massive stars can trigger EO's. In the six cases that have been reported, all of them are related to high massive star formation, the characteristic gaseous filaments are produced in a region of the order of 100 au, and, in some cases, there are runaway massive stars, as in Orion BN/KL. By now, not all EOs are associated to a particular massive star, while a few attempts exist to detect them. However, they are related to maser emission and ultra compact HII (UCHII) regions, contributing to the idea that EOs are produced by massive forming stars \citep{Zapata2017,Zapata2020,Zapata2023}. }

The prototypical EO is found in the Orion~BN/KL region, located at a distance of 400~pc within the Orion molecular cloud, and is the closest example discovered so far \citep{AllenBurton1993,Zapata2009,Bally2017,McCaughrean2023}. It comprises more than two hundred filaments, also known as fingers, expanding quasi-isotropically from a common center, whose kinematics have been studied in infrared through the proper motions of their fingertips. \citep{Bally2015}.
 It has also been found that the mass of gas accelerated by the outflow is about 8 \msun, the longest fingers measuring up to $4\times10^4$ au, with a mean velocity of 30 km s$^{-1}$, and the energy associated with this outflow is $\sim 10^{47}$ erg \citep{Bally2024}. This gives the average mass of each finger on a planetary scale. \citet{Rivera2019a} found that the dynamic model of the plasmon describes the motion of each finger, and managed to determine their initial ejection conditions, such as mass, density, and velocity, and their lifetime. Assuming that they were all ejected simultaneously 500 years ago, \citet{Rivera2019b} have determined that only one-tenth of the fingers would live for more than 1500 years. %would have, determining that only one-tenth of the fingers would live for more than 1500 years. 

One of the most important characteristics of Orion's explosion is the presence of a few runaway massive protostars receding from a common center coincident with the EO's origin \citep{Rodriguez2005,Bally2020}. This has been interpreted as evidence of a violent disintegration of a pre-existing multiple system containing a few protostars of intermediate-to-high masses, but it is not clear how the fingers are produced in such interaction, given their high velocity compared with the runaway stars. 

In this article, 
we propose the possibility that a 
runaway star may have been able to destroy a 
prestellar cluster of particles, resembling very dense clumps, as a first approximation of the dense and compact fingertips producing the Orion BN/KL EO. 
That is, we consider the interaction of a stellar object that 'collides' with a clump particle cluster with a total mass of a stellar object, similar to the protostellar disk GGD27 \citep{Yamamuro2023}. Then, the ejected particles would interact with the dense surrounding medium, leaving long gas wakes behind. This kind of interaction had been analyzed previously by \citet[hereafter Paper I]{Rivera2021}, but in the ideal case of a massless cluster, and by \citet{Arunima2023} in the case of colliding stellar clusters,   or by \citet{Cournoyer24} in the case of the ejection of stars from a hierarchical cluster. The result of such interaction powers the ejection of the cluster members with explosive characteristics. 
In Sec \ref{sec:sec2} we propose a simple analysis of the energy transferred from the moving object to the cluster assuming momentum conservation to describe the analytical frame to study the gravitational interactions. Then, in Sec. \ref{sec:sec3}, we describe the $N$-body numerical simulations, their initial conditions, and the parameters used in each model. In \ref{sec:sec4},  we compare the analytical frame with the results obtained from the numerical simulations. After that, in Sec. \ref{sec:sec5} we consider the implications of this ejection mechanism to explain the existence of explosive outflows. Finally, we summarize our results in Sec.  \ref{sec:sec6}.

\section{The cluster-particle collision}
\label{sec:sec2}

We can start establishing a spherical cluster with radius $r_c$ formed by $N_t$ identical particles whose individual mass is $m_i$, and therefore, total mass $m_c=N_t m_i$.  Also, the distribution of particles is such that its numerical density $n$ is a constant. Then, considering the particle size is much shorter than their mean {particle distance} $n^{-1/3}$, and that these particles are in gravitational balance, orbiting circularly around the cluster center. Because of the homogeneous particle distribution, the mass $m(r)$ enclosed in a radius $r$ from the cluster center is an increasing function of $r$, $m(r)=m_c (r/r_c)^3$ and, therefore, the orbital velocity of a particle at position $r$ is 

\begin{equation}
    v=v_o(r/r_c),
    \label{eq:vo}
\end{equation}

defining $v_o=(m_c G/r_c)^{1/2}$ as the orbital velocity of the particle at position $r_c$, where $G$ is the universal gravitation constant. This cluster is an autogravitating sphere whose total energy is the sum of its potential and kinetic energies

\begin{equation}
E_c= -\frac{3}{10}\frac{Gm_c^2}{r_c},  
\label{eq:Ec}
\end{equation}
{ which considers the cluster as a uniform and continuous distribution of mass in virial equilibrium}.
At the same time, a particle with mass $m_*$, travels from infinity with velocity $v_{*0}$ in the direction of the cluster center of mass, { that is, with an impact parameter of zero}, that we will call {\it bullet} hereafter. The bullet mass $m_*$ is larger than $m_i$,  and $v_{*0}$ is also larger than $v$, which implies that the cluster particles could be considered static when the distance between the cluster and the bullet is only a few times $r_c$. 

\subsection{Momentum and energy conservation}

Since the bullet and the cluster are a closed system, their interaction can be considered a collision, and, therefore, it is possible to analyze the momentum transfer between them as a conservative collision. {  It must be clarified that we define a collision as the abrupt perturbation of the initial conditions of the cluster, therefore, we expect that the cluster-bullet collision is equivalent to the $N_T$ interactions of the bullet with each cluster member.} Initially, since the cluster is in equilibrium and its center is considered at rest, it has total momentum zero in any direction, while the bullet contributes with momentum $p_0=m_*v_{*0}$ in the direction of motion ($x$). After the collision, the remaining momentum carried by the bullet, $m_* v_*$, and the cluster $m_c v_c$, which can be scaled as $p_b=m_* v_*/p_0$ and $p_c=m_c v_c/p_0$, {where $v_c$ is the cluster center of mass velocity} {  then momentum conservation can be rewritten as} 

\begin{equation}
    p_b+p_c=1.
    \end{equation}

In the limit where the mass of the cluster is negligible, we can assume that $p_b\sim 1$, which will decrease as  $m_c$ gets larger up to the limit where  $m_c\gg m_b$ and we can consider the bullet as part of the cluster. Then, the bullet momentum would get negligible compared to the momentum obtained by the cluster and therefore will tend asymptotically to $p_b\sim 0$. Given this behavior, we propose an analytic exponential function that will tend to these limits which depends only on the ratio $\mu=m_c/m_*$ and then we can express the momenta as:

\begin{align}
    p_b&= e^{-\alpha \mu} \label{eq:p*}\\
    p_c&=1-e^{-\alpha \mu} , \label{eq:pc}
\end{align}

where $\alpha$ is assumed as a constant. Then, $p_c$ and $p_b$, can be used to find the velocities $u_c=v_c/v_{*0}$ and $u_*=v_*/v_{*0}$ that can be expressed as

\begin{align}
        u_*=&e^{-\alpha \mu}
    \label{eq:u*}\\
    u_c=&\frac{1}{\mu}\left[ 1 -e^{-\alpha \mu} \right]. \label{eq:uc}
\end{align}
{These equations, in the limit when $\mu \to 0$, tell us that $u_*=1$ and $u_c=\alpha$, that can be interpreted as, after the collision, the bullet remains unperturbed and, at the same time, the cluster gets a considerable velocity, since $\alpha=0.61$ (see Appendix). }

It is expected that a fraction of the bullet kinetic energy $E_{b0}=m_*v_{*0}^2/2$ will be transferred to the cluster. Since it is a closed system, mechanical energy is also preserved. Then, it is possible to equate the energy before and after the collision:

\begin{equation}
    \frac{1}{2}m_* v_{*0}^2+E_c=E_{abs}+\frac{1}{2}m_* v_*^2+\frac{1}{2}m_c v_c^2,
\end{equation}

where $E_{abs}$ is the {internal} energy carried by the system. Dividing the last equation by $E_{b0}$ and  defining $\epsilon_{abs}=E_{abs}/E_{b0}$ and $\lambda=-E_c/E_{b0}$, which assures that $\lambda$ is going to be always positive, we can rewrite it as follows.

\begin{equation}
\epsilon_{abs}=1-u_*^2-\mu u_c^2-\lambda.
\label{eq:eabs}
\end{equation}

As assumed, $u_c$ and $u_*$ depend only on the ratio $\mu$ considering our proposed model. Nevertheless, $\lambda$ depends on the initial conditions of the cluster and the bullet. To understand the behavior of the $\epsilon_{abs}$ function, we analyze separately the terms $1-u_*^2-\mu u_c^2$ which go asymptotically from 0 to 1 as $\mu$ increases from 0 to the limit $\mu \to \infty$, and these terms belong to the interval $[0,1]$. This means that the sign of $\epsilon_{abs}$ is then defined by $\lambda$, then, if $E_c/ E_{b0}<1$, the total energy of the system will be positive, and if $E_c/ E_{b0} > 1$, $\epsilon_{abs}$ will be negative, {  since the sign of lambda depends directly on the sign of  the initial total energy $E_c + E_{b0}$ }. Therefore, there is a wide range of values where the cluster can 'resist' the bullet impact. The cluster analysis we performed needs to be compared with the numerical simulations described in Sec \ref{sec:sec3}, but it is not enough to determine the explosive nature of the bullet-cluster encounter. Then, we analyze the energy transfer for each cluster particle.

\subsection{Energy per particle}

Given a homogeneous particle distribution, it is possible to find the energy per particle to get a value of the energy necessary to unbound a particle from the cluster {  \citep{Goldstein2002}}. We can define an energy scale $E_{i0}=Gm_cm_i/r_c$ and now we can estimate the energy per particle, beginning with the potential energy produced by the mass inside a sphere of radius $r<r_c$ to a particle in that position as ${E_{pi}=-\int F\cdot dr=E_{i0}{r_c^{-2}}\int r dr=\frac{1}{2}E_{i0}\left(\frac{r}{r_c}\right)^2+C,}$ where $C$ is an integration constant that can be determined by the continuity condition given at the border of the cluster, $E_{pi}(r_c)=-E_{i0}$, that implies that $C=-(3/2)E_{i0}$ and then, the potential energy per particle is

\begin{equation}
    E_{pi}=\frac{1}{2}\frac{Gm_cm_i}{r_c}\left[ \left( \frac{r}{r_c} \right)^2-3 \right]=\frac{1}{2}E_{i0}\left[ \left( \frac{r}{r_c} \right)^2-3 \right],
    \label{eq:Epi}
\end{equation}

while the kinetic energy per particle is 

\begin{equation}
    E_{ki}=\frac{1}{2}m_i v^2=\frac{1}{2}m_i v_c^2 \left( \frac{r}{r_c} \right)^2.
    \label{eq:Eki}
\end{equation}

Now we can calculate the total energy per particle by adding Equations \ref{eq:Epi} and \ref{eq:Eki}

\begin{equation}
    E_i=E_{pi}+E_{ki}=E_{i0}\left[ \left( \frac{r}{r_c} \right)^2-\frac{3}{2} \right],
    \label{eq:Ei}
\end{equation}

This energy function is valid for the interval $E_i(0)=-3E_{i0}/2$ to $E_i(r_c)=-E_{i0}/2$, which corresponds to the energies at the center and at the border of the cluster, respectively.  {It is remarkable to note that the total potential energy, $ -\frac{3}{5} Gm_c^2/r_c $, of the system is equivalent to the expression in Eq.~\ref{eq:Epi}, assuming a continuous mass distribution. In this case, we can interpret $ E_{pi} \to dE_{pi} $, $ m_i \to dm $, and $ m_c \to m(r) $ in Eq.~\ref{eq:Epi}. Since each mass element is counted twice, (once as a shell of thickness $ dr $ and once as part of the enclosed mass) it follows that  $
E_p = \frac{1}{2} \int dE_{pi}$, inside the cluster.}

Since the energy per particle is a function of the cluster radius, we can find the functions associated with the distribution of particles $f(r)$ and energies $g(E_i)$, defining the number of particles $dN$ in the interval of distances between $r$ and $r+dr$, or between energies $E_i$ and $E_i+dE_i$ as

\begin{equation}
    dN=N_t f(r) dr= N_t g(E_i) dE_i.
    \label{eq:dNa}
\end{equation}

Given that the particle density, {the number of particles inside a radius $r$ is} $N(r)=N_t (r/r_c)^3$, it is possible to differentiate to obtain that  $ f(r)={3r^2}/{r_c^3},$ and, using Equation \ref{eq:dNa} it is clear that $g(E)=f(r) (dE/dr)^{-1}$, which combined with the derivative $(dE/dr)$ obtained from Equation \ref{eq:Ei}, gives 

\begin{equation}
    g(E_i)
    =\frac{3}{2}\left( \frac{G m_c m_i}{r_c} \right)^{-1} \left( \frac{r}{r_c} \right)=\frac{3}{2 E_{i0}} \left(\frac{E_i}{E_{i0}}+\frac{3}{2} \right)^{1/2}.
\label{eq:histogramEnergy}
\end{equation}

This energy distribution function is valid for the interval $E_i(0)=-3E_{i0}/2$ to $E_i(r_c)=-E_{i0}/2$, which corresponds to the energies at the center and at the border of the cluster, respectively and it is useful to determine the number of particles that have energy in an interval $E_i$ and $E_i +dE_i$.

\section{Simulations}
\label{sec:sec3}

{We have performed several $N$-body
simulations of a particle of $m_*=10$ M$_\odot$ ``colliding" with a cluster of $N_T$ small mass particles to reproduce the velocity and energy distributions derived by the encounter, and we compared these simulations with the analytical model presented in the previous section}. The free parameter of the simulations is the total mass of the cluster.

{   
The numerical 
method is a 
symmetrized leapfrog integrator with a variable timestep formalism, which is second order accurate and is able to preserve energy. For the $N-$body solution, we { have considered} $N$ particles with masses $m_i$, and position given by $x_i$, $y_i$ and $z_i$. 
The force between a pair of particles produces an acceleration, and the new position of each of the particles is strongly dependent of the time step $\Delta t$. A very large time step would solve incorrect trajectories and a small time step reproduces the real trajectory of each particle, dramatically increasing the computation time. In order to have an appropriate time step, we used a time step as, 
\begin{equation}
    \Delta t=A*\sqrt{\frac{R_{min}}{a_{max}}} 
    \label{eq:dt}
\end{equation}
where, $R_{min}$ is the mean distance between a pair of particles, and a$_{max}$, is the maximum acceleration of a single particle, $\sqrt{R_{min}/a_{max}}$ is a gravitational timescale, and $A$ is a factor  between $0 \, \to \, 1$ used for stability and precision \citep{Boekholt15}. A stability analysis of this $N$-body code was presented in Paper I \citep{Rivera2021}. To compare the results that we have obtained in Paper I, we selected $r_c=2.67$ au as the cluster radius, whose origin may be a condensation or an overdensity of the envelope of a multiple system, within which the $m_*$ particle is forming. However, a larger size would represent a weaker interaction, where the internal forces should be considered. The total particles are $N_T=200$, {with a typical separation of approximately} $0.11$~au. {This number {ensures} the cluster can be compared with a continuum mass distribution and is {of} the order of the number of filaments observed in Orion BN/KL}. The cluster is initially centered at the origin in the simulations, and we consider a spatial distribution (the number of {particles} per
unit volume) of the form,
  $  N(r) ={3N_T} {(4 \pi r^{3}_c)}^{-1}$,
 for a homogeneous distribution. 

{For simplicity, this cluster is assumed to be in virial equilibrium and we have chosen every particle with a circular orbit around the distribution center of mass,} the orbital velocity of a particle at a radius $r$ is defined by Equation \ref{eq:vo}. We use a random number generator that chooses a number $\eta$ uniformly distributed in the interval [0,1]. The value is
related to the radial distance as $    r=r_c \eta^{1/3},$ from which we can sample $r$ as a function of the random number $\eta$ (for more details, see \citealt{RGETAL07}). We assigned random directions to the position vector of each particle, and velocity vectors corresponding to circular orbits.

We also included a single massive particle with a mass of 10~M$_\odot$, moving towards the particle distributions with $v_0=$100~km/s. At $t=0$, the massive particle starts moving from the Cartesian point (-10 au, 0,0), in a direction parallel to the {\it x-axis} towards the clump distribution. We have developed sets of 10 random distributions in each of our models to obtain a statistically significant result to compare with the velocity distribution. All simulations run for an evolutionary time of 5 yr, considering that it is enough time for the system to consider the impact is over.   Also, we have considered this evolutionary time short enough to ignore hydrodynamic effects, since it is less than a hundredth {of} the estimated age for Orion~BN/KL, {which also contributes to ignore the effect of the potential produced by the parental cloud.}

%\newpage
\section{Results}
\label{sec:sec4}
%To start with \cite{Arce2008} said that

\begin{figure}
    \centering
    \includegraphics[width=0.48\textwidth]{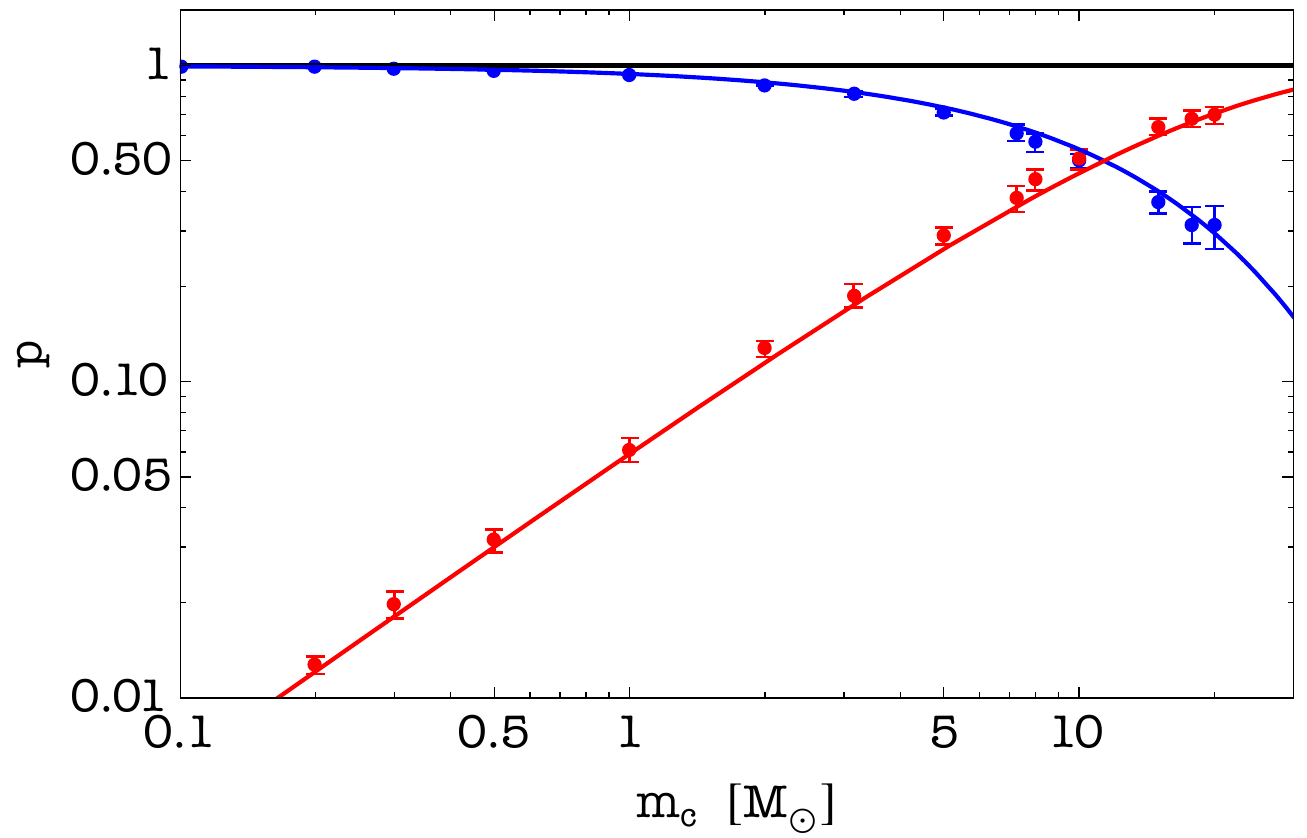}
    \caption{Normalized momentum 5 years after the interaction for each model. The blue points represent the momentum obtained by the star, and the red points correspond to the cluster center of mass velocity. The blue and red lines represent Equations \ref{eq:p*} and \ref{eq:pc}, respectively, {using $\alpha=0.61$}.}
    \label{fig:F1}
\end{figure}

Our first step is to check for the validity of the relations obtained in Section \ref{sec:sec2} for the momentum and velocity obtained by the cluster and the bullet after the impact in Equations (\ref{eq:p*}-\ref{eq:uc}). As described in the last Section, each model has 10 clones. Each one is used to obtain the momentum and velocity for both the bullet and the cluster. To obtain a statistically significant result, we have taken the mean value {of} the 10 clones as its representative value and their standard deviation as the uncertainty {of}  
 the mean value. Then the simulations {yield the} data depicted in Figure \ref{fig:F1}, presenting the normalized momentum for the star particle as blue points and the center of mass of the cluster as red points. We also present the proposed functions for the cluster momentum and, subsequently, for the momentum of the star. Interestingly, the cluster and bullet momenta are the same when their masses are equal and the exponential function from Equation \ref{eq:p*} appears to be an appropriate description for the momentum transfer to the cluster.

\begin{figure}
    \centering
    \includegraphics[width=0.5\textwidth]{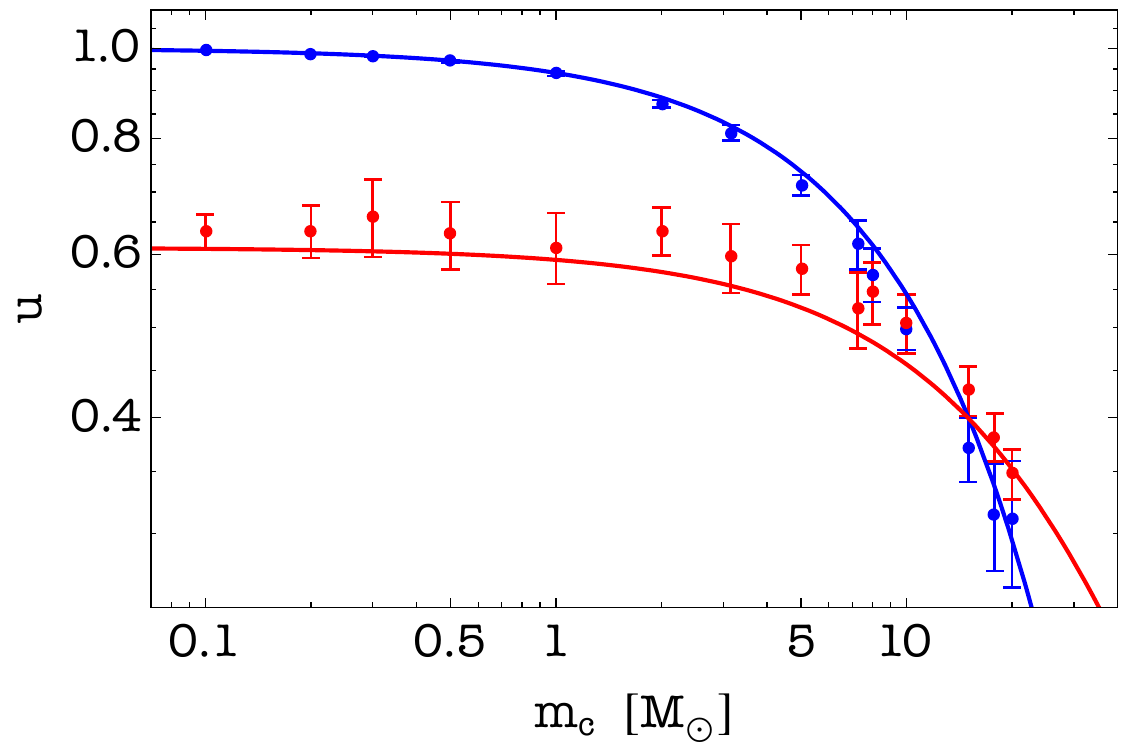}
    \caption{Normalized velocity 5 years after the interaction for each model. The blue points represent the velocity obtained by the star, and the red points correspond to the cluster center of mass velocity. The blue and red lines represent equations \ref{eq:u*} and \ref{eq:uc}, respectively, {using $\alpha=0.61$}.}
    \label{fig:F2}
\end{figure}

Then, in Figure \ref{fig:F2}, we present the velocity of the bullet in blue points and the velocity of the center of mass of the cluster in red points. Also, the velocity functions are represented by solid lines, blue and red, for $u_*$ and $u_c$, respectively. It is outstanding that for low masses, that is $m_c < 1$ M$_\odot$, the effect of the cluster on the star is almost negligible, and the cluster velocity is indeed very close to $v_c\sim 0.61 v_{*0}$, {which coincides with the limit of Eq. \ref{eq:uc} when $\mu\to0$}. Then, at 10 \msun, the cluster and bullet velocities are very similar.

Now, to compare the energy carried by the cluster after the bullet impact, {we obtain an adequate expression for the parameter $\lambda$ in Equation \ref{eq:eabs}, that is

\begin{equation}
    \lambda=-\frac{E_c}{E_{b0}}=\frac{3m_c^2G}{5r_c} \cdot  \frac{1}{m_*v_{*0}^2} ,
\end{equation}
that can be rewritten in convenient units as}
\begin{equation}
    \lambda=5.336\times 10^{-2} \frac{\left(\frac{m_c}{{\rm M}_\odot} \right)^2  }{\left(\frac{r_c}{\rm{au}} \right) \left( \frac{m_*}{{\rm M}_\odot} \right)\left(\frac{v_{*0}}{100~\rm{km\, s}^{-1}} \right)^2}.
\end{equation}

Using the same parameters used in our simulations, $m_*=$10~M$_\odot$, $r_c=2.67$~au and $v_{*0}=100$km s$^{-1}$, we have that

\begin{equation}
    \lambda=1.99\times 10^{-3} \left( \frac{m_c}{{\rm M}_\odot} \right)^2.
\end{equation}

Additionally, using our assumption for the after-impact velocities given by Eqs. \ref{eq:uc} and \ref{eq:u*} we can transform Eq. \ref{eq:eabs} into 

\begin{equation}
    \epsilon_{abs}=1-\lambda 
    -e^{-2\alpha \mu }
    -\frac{1}{\mu}
    \left[ 1- e^{-\alpha \mu }\right]^2
    \label{eq:epsiloni}
\end{equation}

\begin{figure}
    \centering
    \includegraphics[width=0.45\textwidth,trim=0 15mm 0 15mm,clip=true]{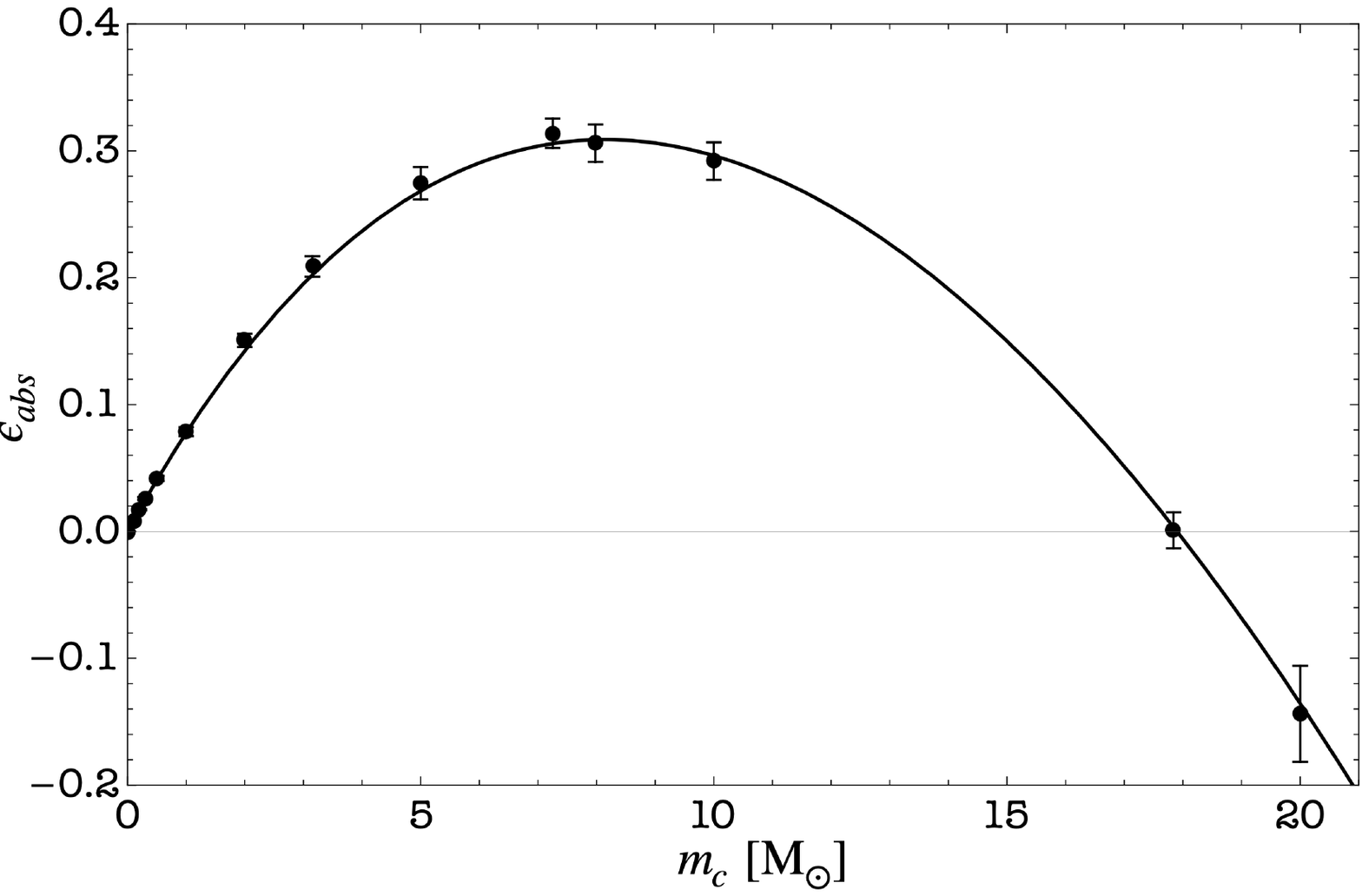}
    \caption{Internal energy $\mathbf{\epsilon_{abs}}$ after the collision for a cluster with a mass $m_c$. The points represent the internal energy calculated in the simulations, while the solid line shows the expected behavior of the energy from Equation \ref{eq:epsiloni}.  This energy is positive up to 17.84 \msun. }
    \label{fig:F3}
\end{figure}

According to Eq. \ref{eq:epsiloni}, $\epsilon_{abs}$ is positive for low mass clusters and negative for high mass clusters, being  $m_c =$ 17.84 M$_\odot$ the limit where $\epsilon_{abs}$ changes signs. {  This is expected since, as the cluster {becomes} more massive, the bullet energy gets smaller compared with the cluster internal energy and, therefore, the cluster does not get affected significantly.}
%a massive cluster tends to be resistive to the impact of a disruptive particle by its own inertia‘}.
In Figure \ref{fig:F3} we show the internal energy $\epsilon_{abs}$ for each model after the collision, taking the mean value, plotted by black points, its dispersion, obtained by the 10 clone models as the uncertainty, and a solid line as the analytical model given by Eq. \ref{eq:epsiloni}.  Also, we can interpret that when the total internal energy $\epsilon_{abs}$ is negative, the star is also bound to the cluster, that is, it {is} absorbed and trapped by the cluster.

\subsection{The {unbound {population}}}
\label{sec:particle}

% \begin{figure}[!h]
%     \centering    
%     \includegraphics[width=0.4\textwidth]{EvsR.pdf}
%     \caption{Energy per cluster member as a function of its radius $r$, in the case of the analytic model (solid line), and the models with $m_c=0.1$M$_\odot$ (dashed line), and $m_c=10$~M$_\odot$ (dot-dashed line)}
%     \label{fig:F4}
% \end{figure} 

\begin{figure}
    \centering    \includegraphics[width=0.45\textwidth]{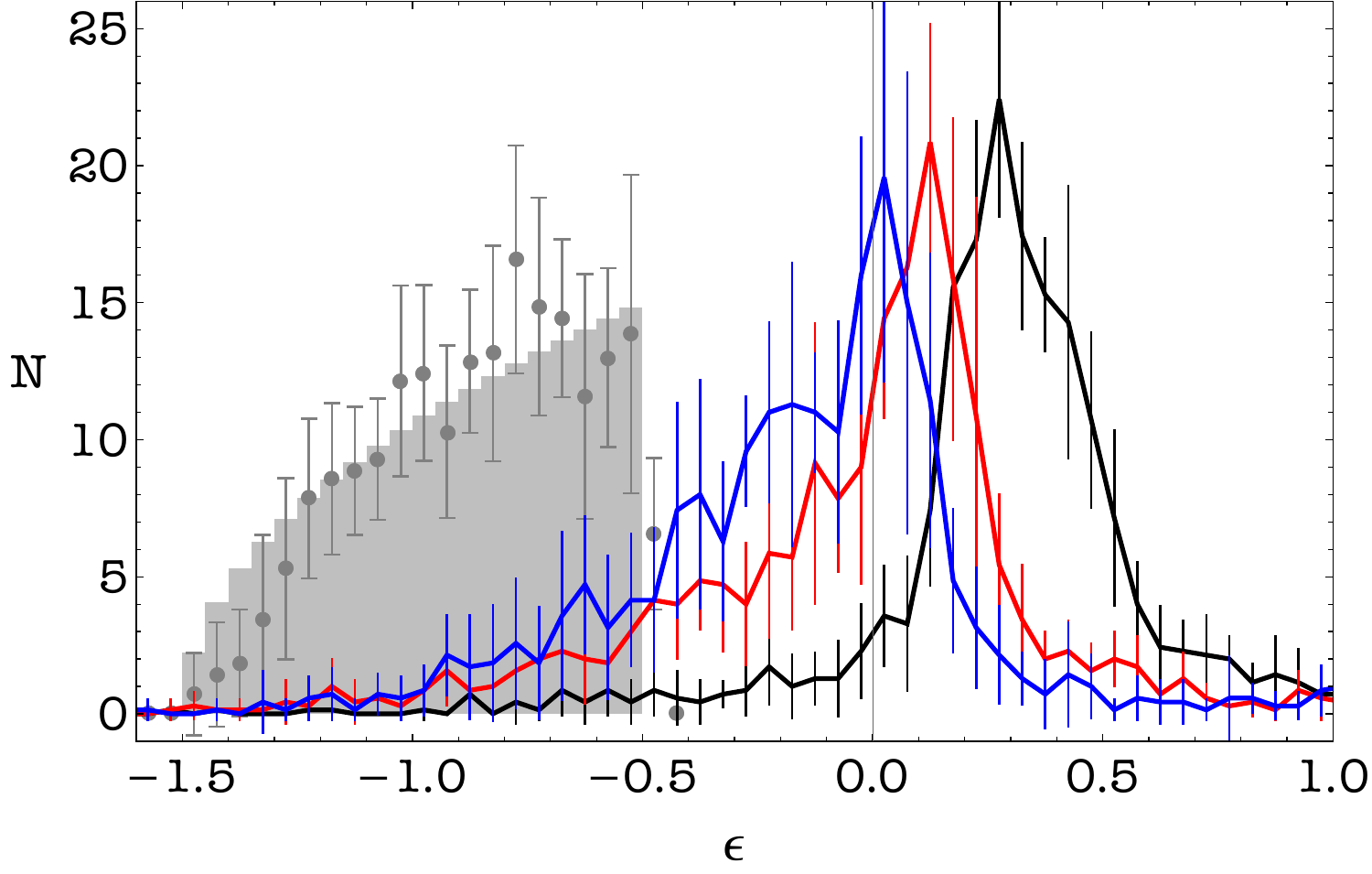}
    \caption{Energy $\epsilon$ histograms before the collision for  the model $m_c=10$M$_\odot$~ (gray points), compared with the analytic model (gray bars) and after the collision for the models 10\msun ~in black, 15 \msun ~in red, and 20 ~\msun in blue. }
    \label{fig:F5}
\end{figure}

In the last Section, we obtained the total internal energy for the impacted cluster. This describes when the bullet gets trapped by the cluster and helps us to obtain a description of the overall velocities obtained by the star and the cluster because of the interaction. Nevertheless, this is still far from giving us a detailed picture of the interaction. To grasp the consequence of the impact {on} the cluster, we need to describe the energy obtained by each particle. To achieve this, in the simulation, we calculated the total energy per particle $\epsilon$, assuming that a particle with energy greater than zero is {unbound} and for $\epsilon \leq 0$ this particle would {still be} {bound} to its neighbors or the bullet. First, we compare the energy distribution for the system before the interaction, which can be done analytically for a cluster in dynamic equilibrium, studying the energy per particle as in Sec. 2, directly measuring the energy per particle in each numerical model.

In the initial configuration, we define the energy per particle inside a homogeneous sphere as

\begin{equation}
  \epsilon(r)=
    -\frac{1}{2} \epsilon_{i,0} \left[ 3- \left (\frac{r}{r_c}\right)^2 \right ],
    \label{eq:energyperparticleVSr}
\end{equation}
using $\epsilon_{i,0}=m_cm_iG/r_c$ as the potential energy of a particle on the surface of the sphere with radius $r_c$. Then, we have checked that every model follows the analytic distribution of energy as a function of the cluster radius, as in Equation \ref{eq:energyperparticleVSr} with an uncertainty of less than 5\%.

\begin{figure}
    \centering
    \includegraphics[width=\columnwidth]{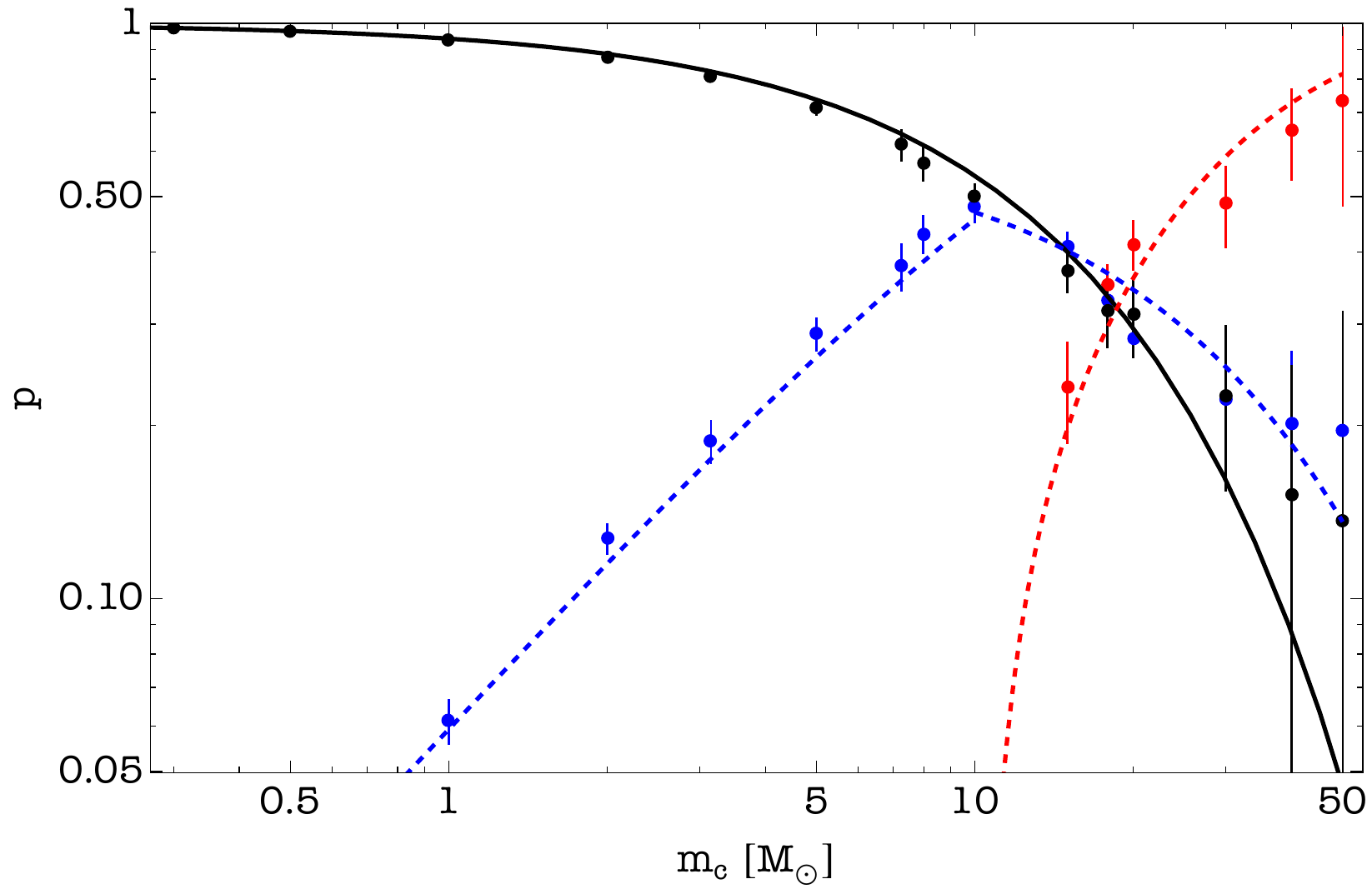}
    \caption{Normalized momentum for bullet (black points), the bound cluster (red points), and the unbound population ({blue} points). The points are the values obtained from the simulations. The black line represents Equation \ref{eq:p*}. The blue line represents Equation \ref{eq:pc} below $m_c=10$\msun~ and $0.64e^{0.31\mu}$ for $m_c>10$\msun. The red line represents $p_{c,u}=1-p_{c,b}-p_*$. }
    \label{fig:F6}
\end{figure}

A useful tool to estimate the energy transfer is to produce an energy histogram as in Figure \ref{fig:F5}, using Equation \ref{eq:histogramEnergy} to find the number of particles in the {energy range} between $-1.5 \epsilon_{i,0}$ and $-.5 \epsilon_{i,0}$ in bins of $0.1 \epsilon_{i,0}$ wide and counting the number of particles in the simulations, using the same energy bins. {The} black bars in \ref{fig:F5} represent the energy histogram in the initial situation for the analytical model, and the black points correspond to numerical simulations. Additionally, in the same Figure are represented, in grey, red and blue, the energy histograms for the 10 \msun, 15 \msun, and 20 \msun ~models, respectively, after the interaction of the bullet with the cluster. These histograms show how particles transitioned from having negative energy to having both negative and positive energy. Negative energies correspond to particles bound together, while positive energies correspond to ejected particles. Therefore, we can say that the cluster, which originally was completely {bound}, is now divided into a {bound} cluster and an {unbound population}. Henceforth, we need to analyze them separately. In summary, the interaction of the bullet with the cluster has led to a significant transformation in the energy distribution. This result suggests the importance of studying these two subsets separately to understand the post-interaction dynamics better. Moving forward, the detailed analysis of these two particle populations will provide crucial insights into the system's evolution following the interaction, offering a more comprehensive understanding of the underlying physical processes.

\begin{figure}
    \centering
    \includegraphics[width=\columnwidth]{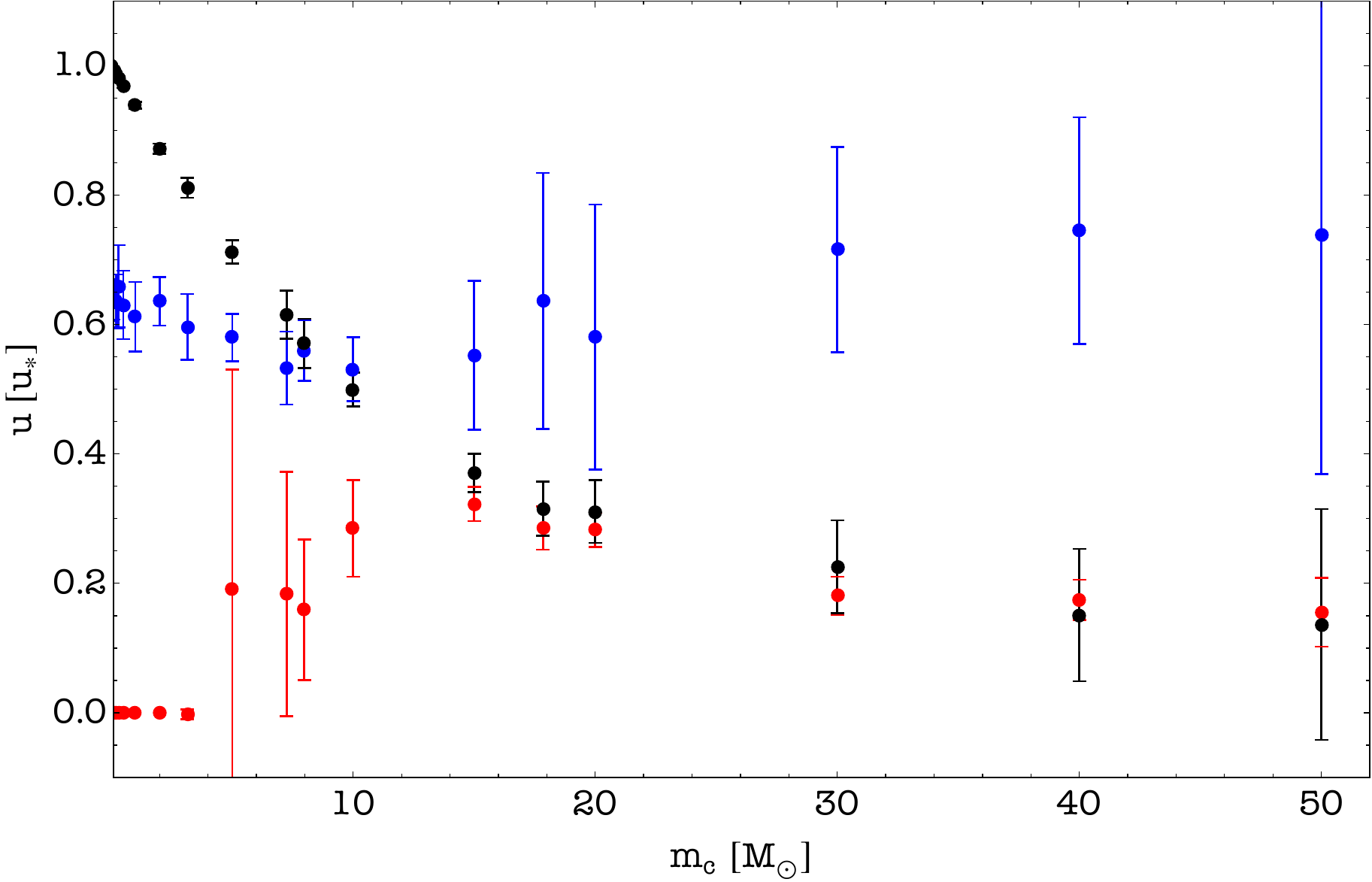}
    \caption{{  Normalized velocity} of the bullet (black points), and averaged velocity of the bound (red) and the unbound (blue) particles of the targeted cluster.}
    \label{fig:F7}
\end{figure}

Figure \ref{fig:F6} is similar to Figure \ref{fig:F1}, displaying the momentum acquired by the bullet $p_*$ (black points), the bound cluster $p_{c,b}$ (red points), and the unbound {population}, $p_{c,u}$ (blue points),  as a function of $m_c$. It can be observed that, as $m_c$ increases, the unbound population momentum also increases until it acquires a momentum comparable to that of the bullet. At the same time, the bound cluster asymptotically absorbs the total momentum until it reaches the total momentum of the system. Simultaneously, this indicates that the momentum acquired by the unbound population, i.e., particles with positive energy, gradually represents a smaller fraction of the total available momentum, and has a tendency to drop to zero, which should be related to the comparison of the internal energy of the original cluster with the bullet energy. The black line represents the fit proposed by Equation \ref{eq:p*}, while the blue line represents a fit similar to Equation \ref{eq:pc} for the momentum carried by the unbound particles for ~$m_c<10$\msun, and $p_{c,u}=0.64e^{-0.3\mu}$  from $m_c=10$\msun, and the red line represents the remaining momentum of the bound particles $p_{c,u}=1-p_{c,b}-p_*$. On the other hand, Figure \ref{fig:F7} shows the mean velocity acquired by each of these components. Black points represent the velocity acquired by the star, the mean velocity acquired by the unbound population is shown in blue points, and the mean velocity acquired by the bound cluster in red points. Once again, it can be observed that the behavior of the released particles indicates that their center of mass moves at an almost constant speed, independently of the analyzed model, that is, the mass of the cluster. Additionally, it can be concluded from Fig. \ref{fig:F7} that the bullet unbinds all particles in clusters with $m_c < 5$M$_\odot$, and in the range $5$M$_\odot \le m_c \le 10$M$_\odot$, there are bound particles, but their momentum is negligible as shown by Fig. \ref{fig:F6}, which indicates only a small fraction of particles remain bound.

\begin{figure*}
    \centering
    \includegraphics[width=\textwidth]{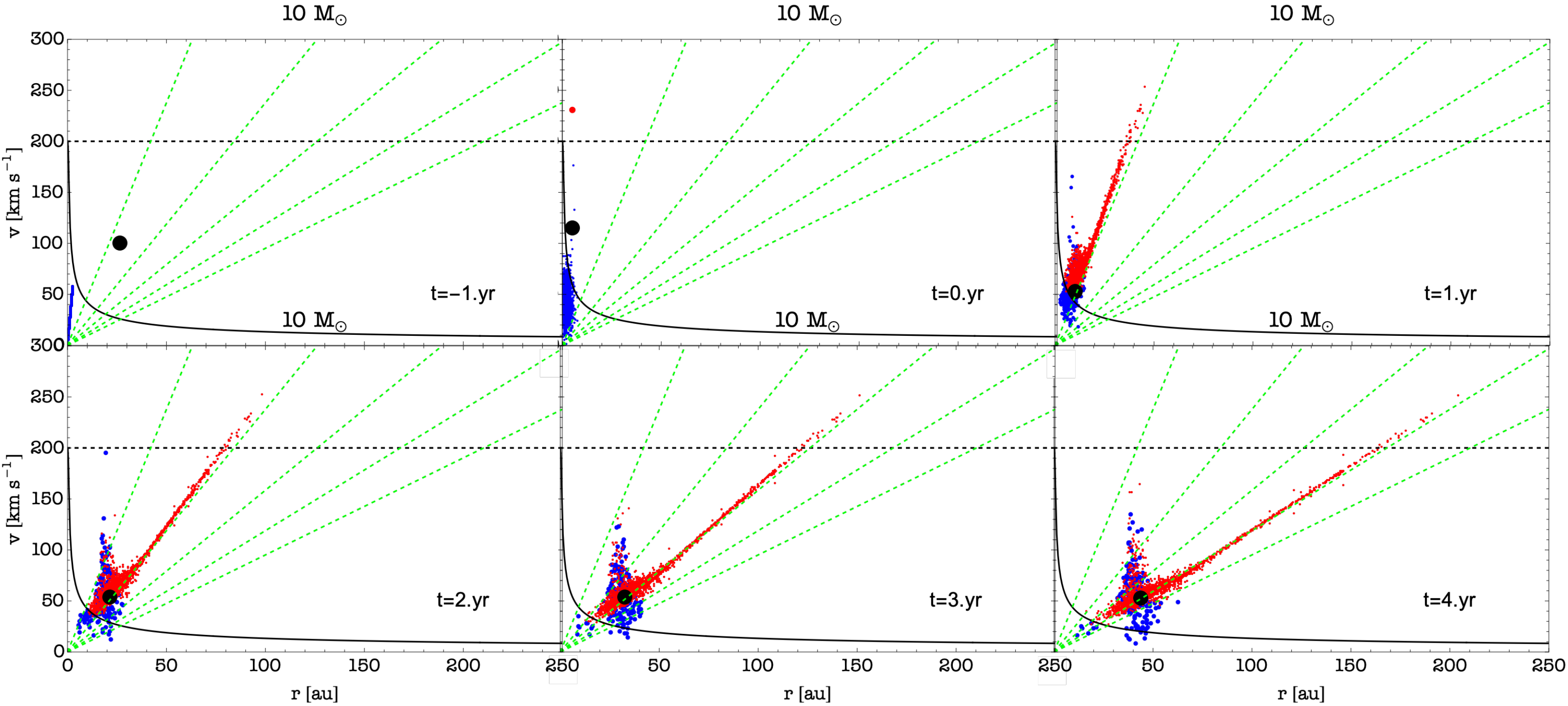}
    \caption{Phase Diagram ($|r|$[au] vs $v$[km/s]) for the model with $m_c=10$~M$_\odot$ at different times. The bullet is represented by the black point and the green dotted lines, from left to right, represent the dynamical ages of 1, 2, 3, 4 and 5 years from the closest encounter, which is considered $t=0$ yr. }
    \label{fig:F8}
\end{figure*}

The panels in Figure \ref{fig:F8} show phase diagrams at different times during the simulation counting from the time when the bullet reaches the closest distance to the cluster, as marked in the lower right corner of each panel. The star is denoted by a black dot, while the bound cluster corresponds to the blue points, and the unbound cluster to the red particles. The horizontal axis corresponds to the absolute distance from the origin, and the black dashed line represents twice the initial velocity of the bullet, while the green lines show the dynamical age from one to five years after the collision from left to right. It can be observed that the bullet is initially one year away from colliding with the cluster. In Figure \ref{fig:F8} (first panel, second row), we can see that the interaction has ended, and the unbound particles lie along a Hubble's line, whose slope corresponds to the dynamical age since the close interaction took place, that is, $t=1$ yr. However, the star remains in the same radial position as the bound cluster, indicating that the bullet seems to have become part of the bound cluster. Also, every panel shows that after the interaction several particles acquire a velocity greater than 200~km~s$^{-1}$ (dashed lines), which was the maximum velocity that a particle could obtain in the massless analytical approximation, {  and twice the bullet velocity of 100~km~s$^{-1}$ \citep{Rivera2021}}. In fact, this limit tends to increase linearly as a function of the cluster mass, as shown in Figure \ref{fig:F14}.  The analytical approximation was obtained for a massless cluster without energy transfer since the cluster mass is negligible. For small values of $m_c$, the maximum velocity coincides with this analytical limit. As $m_c$ increases, the energy transfer becomes efficient, possibly forming close binaries, and ejecting particles faster than 200 km~s$^{-1}$. This result can help to understand the large velocities found in EOs,  and could be related to the compactness of the bound cluster, which would increase the kinetic energy of the unbound population.

\begin{figure}
    \centering
    \includegraphics[width=\columnwidth]{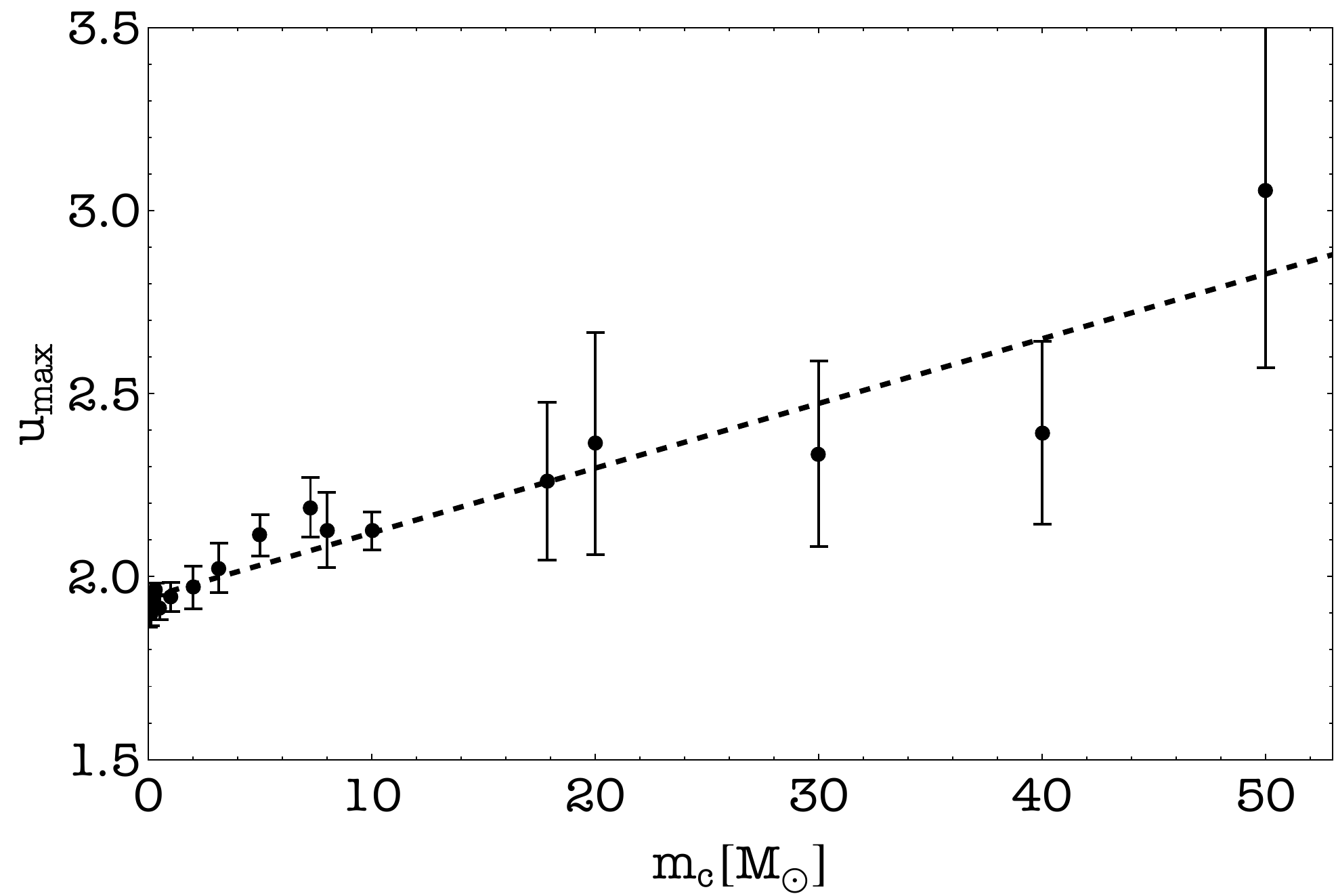}
    \caption{Normalized velocity of the fastest particle ejected after the bullet-cluster collision, in units of the bullet velocity. i. e. 100~km~s$^{-1}$. There is a linear trend, where the normalized maximum velocity tends to increase as the cluster gets more massive, going from a value of 2, in the case of a massless cluster, to $u_{max}\sim 3$, for a cluster with $m_c=50$\msun. The dashed line represents a linear fit to the data.}
    \label{fig:F14}
\end{figure}

\subsection{Evaporation parameter}

Also, we define an evaporation parameter $\beta$, which is the fraction of particles that acquire positive energy and therefore escape the gravitational potential. In our simulations, the evaporation parameter is defined as the fraction of particles that get a velocity larger than the escape velocity and are therefore, unbound, moving in free motion. Figure \ref{fig:F9} presents the ratio of the liberated particles to the total number of particles in the cluster over time. It can be seen that this number is almost constant after one year, indicating that, similarly to Figure \ref{fig:F8}, the interaction is almost instantaneous, and once the evaporation factor is reached, it does not considerably increase over time. Three models for different values of $m_c$ are shown in the figure: 0.1\msun ~in orange, 10 \msun ~in blue, and 17.84 \msun ~in black.

\begin{figure}
    \centering
\includegraphics[width=0.45\textwidth]{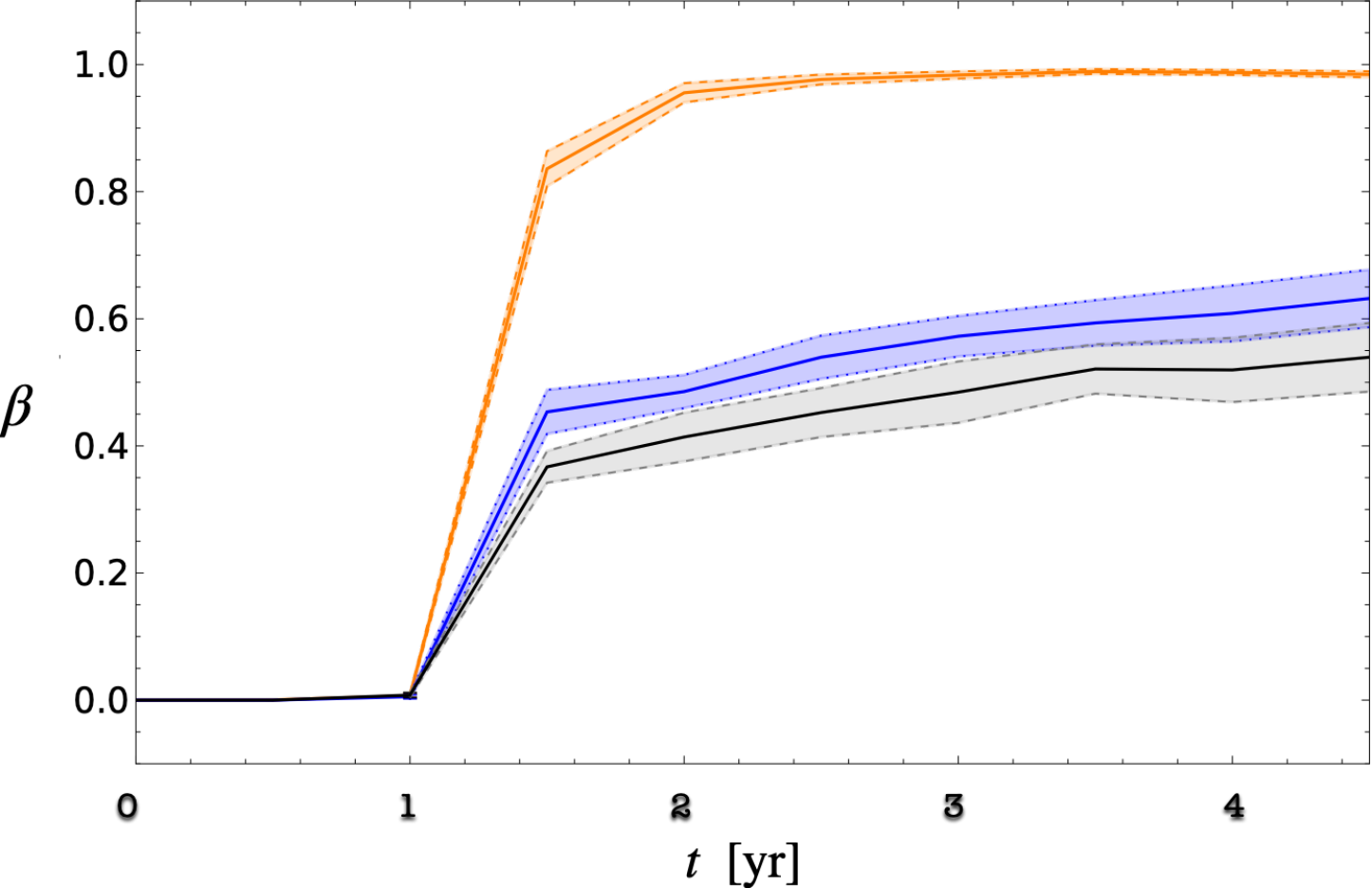}
    \caption{Evolution of the evaporation parameter for the clusters of 0.1\msun ~in orange, 10 \msun ~in blue, and 17.84 \msun ~in black. The shadowed colors along each line represent the uncertainty of the mean value. }
    \label{fig:F9}
\end{figure}

 Another way to understand this interaction is by dividing our cluster into layers, each with the same number of particles, and finding the evaporation factor for each of these layers. This is shown in Figure \ref{fig:F10} for the models of 10 \msun, 15 \msun, 17.84 \msun, and 20 \msun. It can be observed that in each layer, the evaporation factor appears to be constant with a slight positive slope, indicating that the outer layers or shells seem to evaporate slightly more than the inner shells. However, this effect is very small, and more statistics are needed, as apparently, all layers evaporate with a constant factor. The models are represented in red for the 10 \msun ~model, green for the 15 \msun ~model, blue for the 17.84 \msun ~model, and black for the 20 \msun ~model. Solid lines represent the linear fit that best fits the data, and dashed lines are the average value of the evaporation factor for each layer. Also, this figure suggests that our results should stand for a smaller number of particles than $N_t$ since every layer seems to behave independently from the others.

Finally, in Figure \ref{fig:F11}, we plot the total evaporation factor for each of the models, ranging $m_c$ from $10^{-8}$  to 50 \msun. This factor ranges from one when the entire cluster has evaporated to zero when the star is absorbed, and very few particles evaporate. The behavior of this curve is asymptotic for both ends of the fitted point. We have used a four-parameter {arc-tangent function as the most suitable empirical fit.

\begin{figure}
    \centering
    \includegraphics[width=0.5\textwidth]{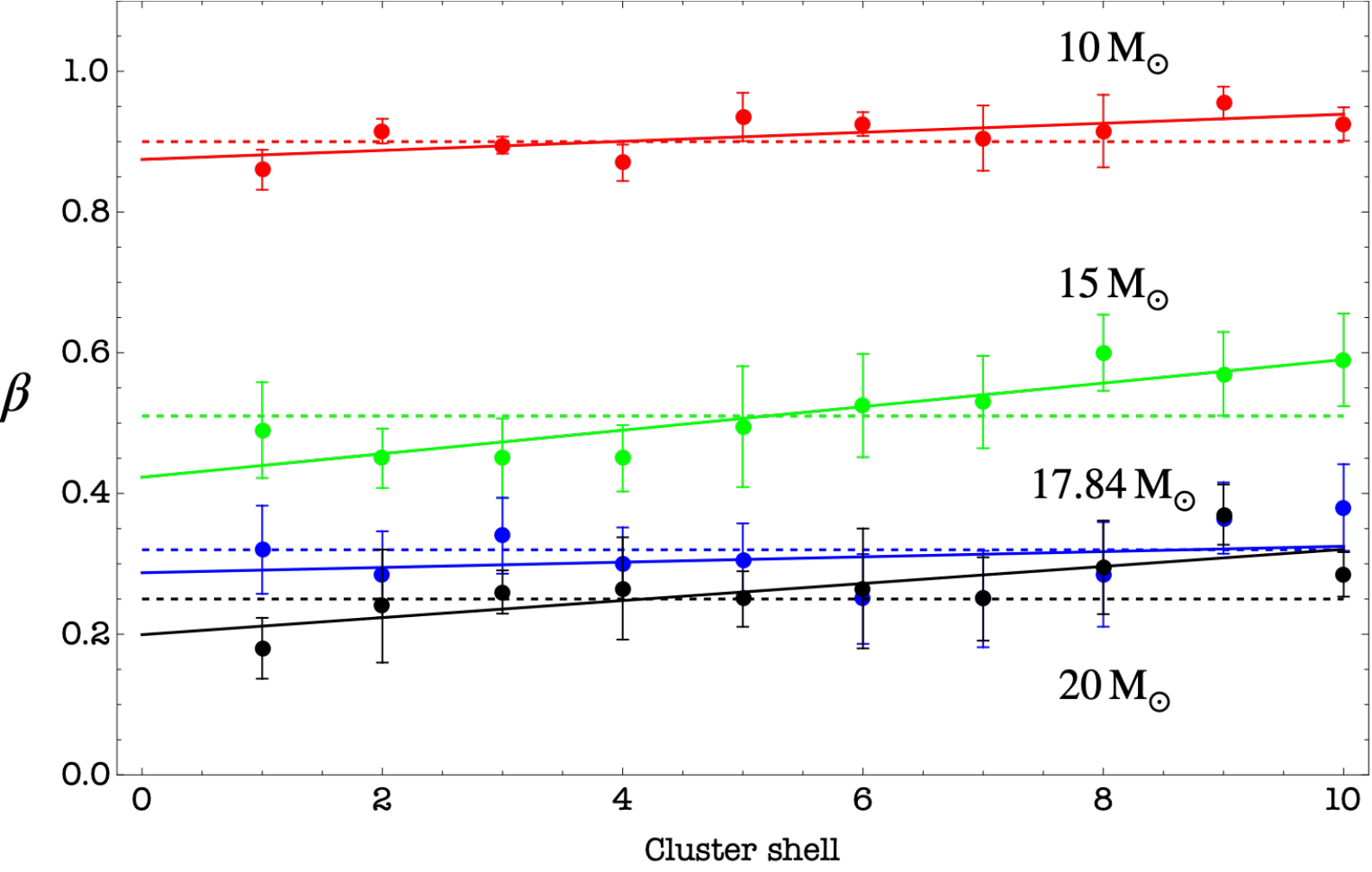}
    \caption{Evaporation factor for every shell represented with points in the model $m_c=10~$M$_\odot$ (in red),
    $m_c=15$~M$_\odot$ (in green),
    $m_c=17.84$~M$_\odot$ (in blue), and
    $m_c=20$~M$_\odot$ (in black). The continuous lines represent the linear fit of each model while the dotted line represents the mean value of the evaporation parameter per shell.}
    \label{fig:F10}
\end{figure}

As shown in Figure \ref{fig:F11}, the evaporation parameter goes from 1 to 0 asymptotically as the mass of the cluster increases, which means that the cluster becomes more resistant to ejecting particles as it is more massive  which must be related to the bullet energy compared with the cluster energy. Empirically, we obtained a fit for the evaporation factor with an arc-tangent function:

\begin{equation}
\beta(m_c)=a\arctan(b m_c +c)+ d,    \label{eq:evaporation}
\end{equation}

with values $a \to -0.33,\, b \to 0.31,\, c \to -4.6,\, d \to 0.57$

\begin{figure}
    \centering
    \includegraphics[width=0.45\textwidth]{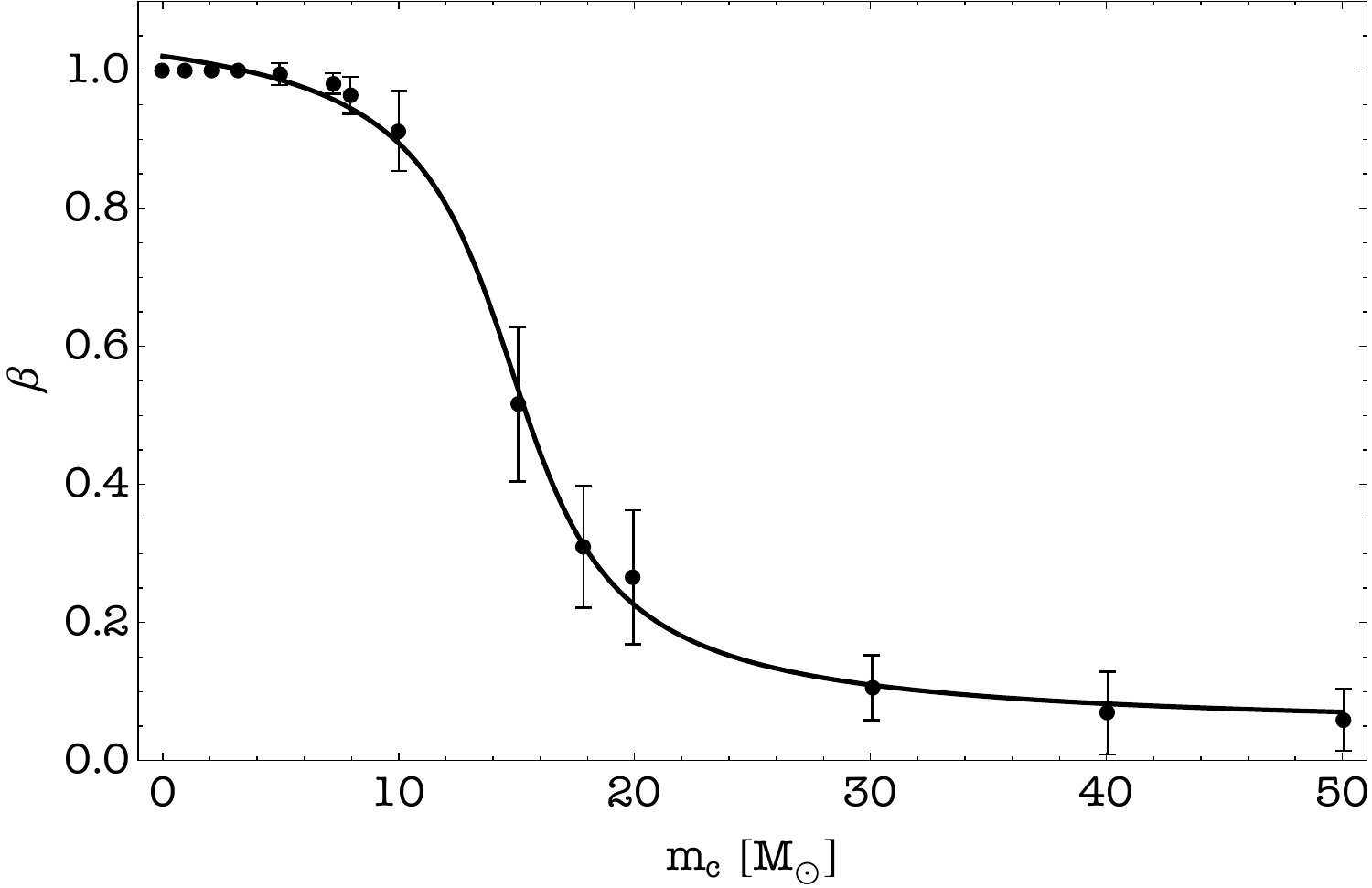}
    \caption{Evaporation parameter $\beta$ as function of the cluster mass $m_c$ {  (black points)}. As expected, a cluster gets evaporated easily as its mass decreases and tends to absorb the bullet as its mass increases. {  The solid line represents an empirical fit explained in the text.}}
    \label{fig:F11}
\end{figure}

\section{Discussion}
\label{sec:sec5}
The mechanism that originated the explosive outflow Orion BN/KL remains unclear. However, our model is a first approximation to explain the mechanism producing EOs.  One of the main problems that emerged in paper I is that the isotropy observed in Orion BN/KL cannot be reproduced. This is because our dynamic explosion model is based on the passage of an object runaway, i.e., a runaway star, which interrupts the balance of a fragmented cloud. However, paper I predicted a preferential direction in the movement of the ejected fragments.
\begin{figure}
    \centering
    \includegraphics[width=0.77\columnwidth,trim={2.cm 0.7cm 1cm 0.5cm},clip=true]{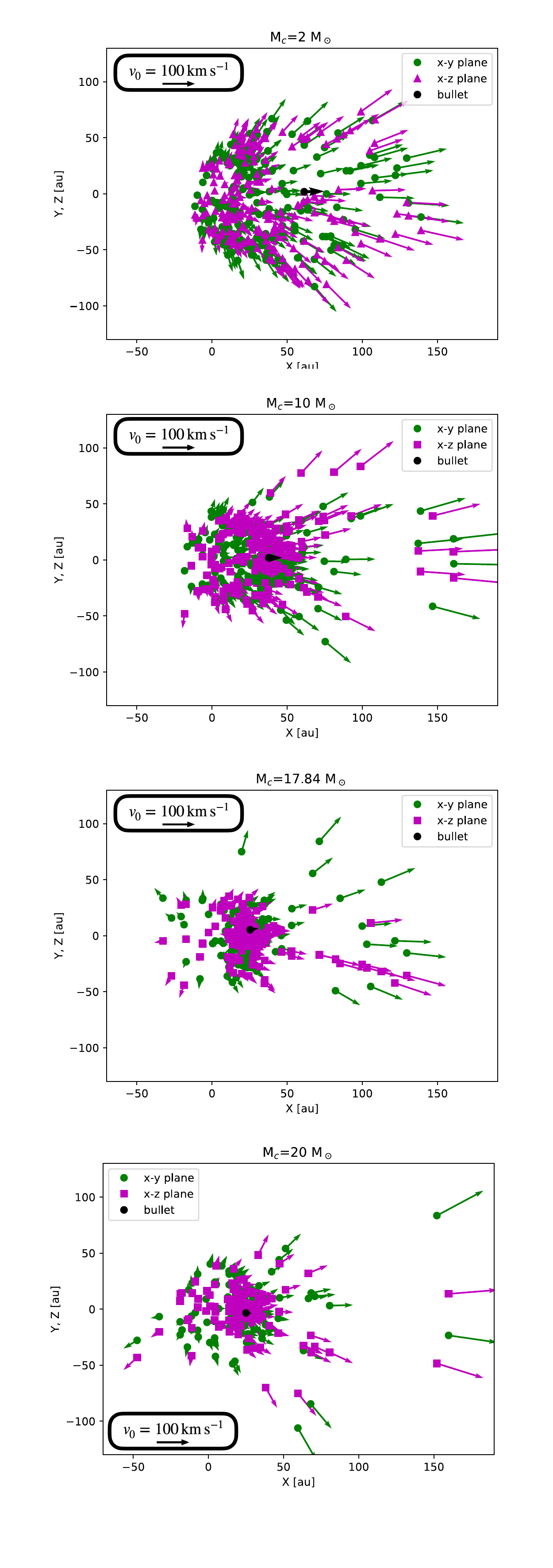}
    \caption{The position of the particles of the model with masses of the cluster of  2, and 10, 17.84 and 20 M$_\odot$, upper, central-upper, central-lower, and lower panels, respectively. The green circles and magenta squares represent the particles traveling in the $xy$ and $xz$ plane, respectively, while their associated arrows represent their velocity. {  The black arrow represents the length of a velocity vector of 100 km s$^{-1}$ and its direction in the $xy$ plane represents the initial bullet direction.} }
    \label{fig:F12}
\end{figure}

As in Paper I, we compare the results obtained by letting a massive particle (10 M$_\odot$) interact with a group of 200 particles with a total mass of 2, and 10, 17.84 and 20 M$_\odot$ (see Figure~\ref{fig:F12}). The green circles and magenta squares represent the point of view from the $xy$ and $xz$ planes, respectively, and their associated arrows quantify their velocity in terms of the bullet velocity $v_*$ that moves in the $xy$, represented with a black arrow in the top-left corner of each panel. As we can see, the less massive models are much more similar to those presented in our paper I, where the mass of the cluster is negligible, and the interaction with the massive particle produces an explosion with a preferential direction, following the trajectory of the massive object. In the upper panel of Figure \ref{fig:F12}, where the mass of the cluster is 2 M$_\odot$, we can observe that certain particles have not gained much velocity, and some of them are moving in directions opposite to the motion of the massive object. On the other hand, in the central-upper panel, when the mass of the cluster is equal to the mass of the bullet, approximately half of the particles composing the cluster are disturbed, moving in a preferential direction along the trajectory of the perturbing object. Another significant portion of the objects do not follow this movement, exhibiting a much more isotropic distribution. In Figure~\ref{fig:F12}, the central-lower panel and the bottom panel (17.84 and 20 M$_\odot$, respectively), show that the velocity distribution no longer predominantly follows the direction of the trajectory of the 10 M$_\odot$ bullet. In addition, numerous small objects remain attached to the object {(black star in Figure~ \ref{fig:F12})} that has disturbed them.

{{Also, the produced explosion is not isotropic from the static frame of reference, since most of the particles move in the direction of the bullet. Nevertheless, it is useful to follow the center of mass of the ejected particles, that is, to change the position and velocity of each of the ejected particles with respect to the position and velocity of the center of mass of this unbound system. In Figure \ref{fig:Fig13}, we explore the geometry of the explosion in the center of mass frame of reference for the model with $m_c=10 \rm{M}_\odot$ where the origin is located at the center of mass. In this frame, the positions of each particle are aligned with their velocities (Figure \ref{fig:Fig13}, upper panel).  Even more, it is possible to obtain a histogram for the number of particles between the coordinates $x_i$ and $x_i +dx$, where $x_i$ is the position along each coordinate axis in this frame of reference. For a symmetric explosion, these histograms are expected to be symmetric independently of the chosen axis (Figure \ref{fig:Fig13}, middle panel). Then, the isotropy can be obtained by counting the number of particles with an ejection direction angle between $\theta$ and $\theta+ d\theta$, where $\theta$ is the counterclockwise angle between the $x-$axis and the position vector of a given particle (Figure \ref{fig:Fig13}, bottom panel). Then, the ejection is isotropic, since there are particles ejected in all directions, even when it is not completely symmetric.}}

The relationship between our dynamical results and Orion BN/KL type events is much more evident when we compare the kinematics of the modeled collision. For example, observations of dynamic explosions show that gas material (particles) further from the origin of the explosion have higher velocities and that this increase in velocity is linear with distance. Even when our model is dealing with ideal particles, we have been able to reproduce the velocities and energies observed in Orion~BN/KL in a scale of 100 au from the interaction point, and these particles would have to interact with the dense surrounding medium, creating the observed fingers. This speculation should be explored with a hydrodynamic code, taking into account the surrounding gas.

In Figure \ref{fig:F15} we present the results of the model with a total cluster mass of 20 M$_\odot$, for 4 evolutionary times 0, 0.5, 2, and 3 yr (top-left, top-right panel, bottom-left and bottom-right, respectively), and there is a linear trend of the ejected particles, while the bound particles travel together with a wide velocity distribution. The continuous line represents the escape velocity required by a particle in the initial configuration to get ejected from the cluster. At a first approximation, particles that can get away far enough from the massive bullet with a velocity larger than the initial escape velocity can be considered part of the EO generated by the bullet-cluster interaction.

\begin{figure}
    \centering
    \includegraphics[width=\linewidth,trim={4cm 0 4cm 0},clip]{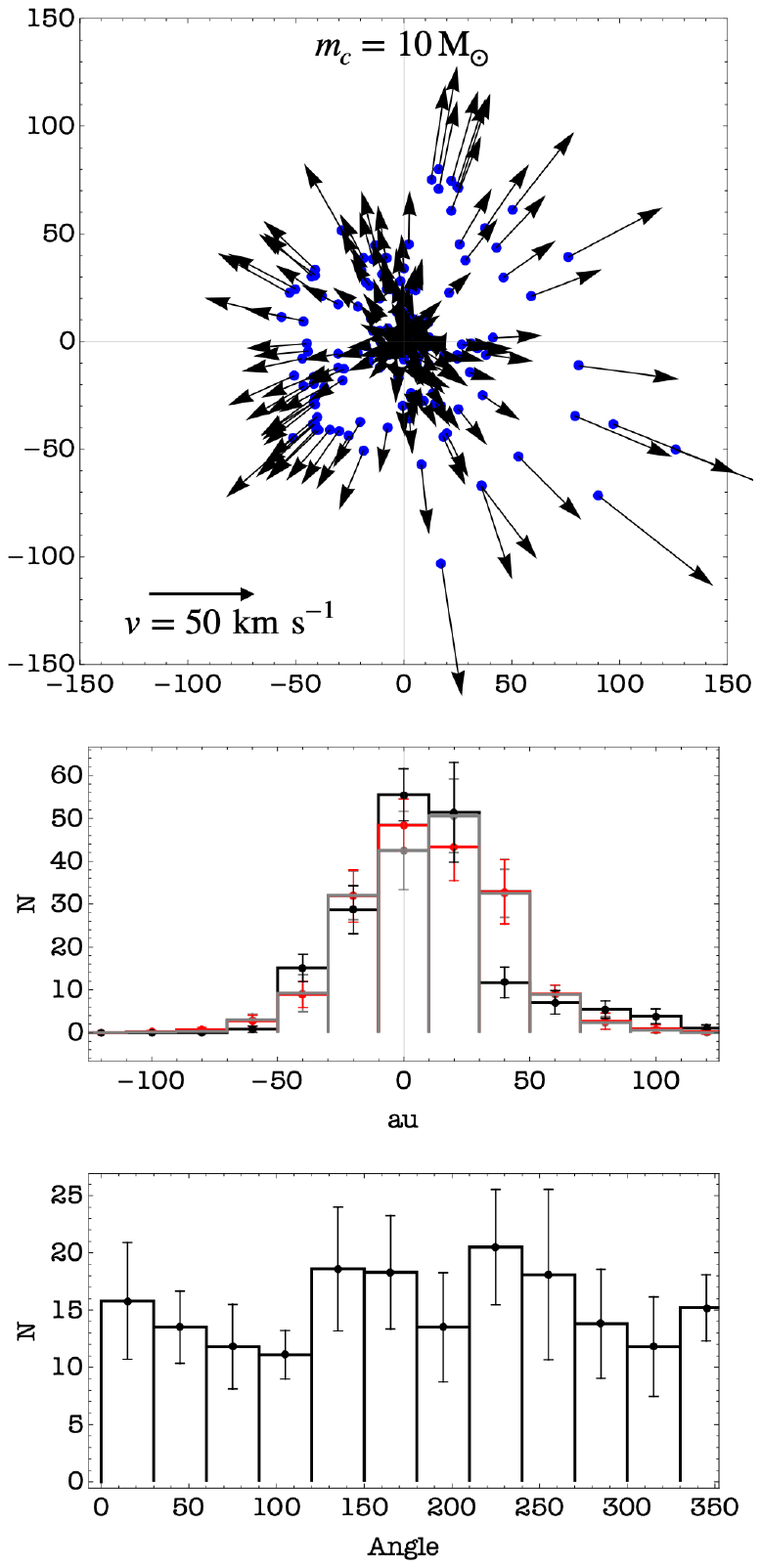}
    \caption{  Explosion isotropy from a frame of reference that follows the center of mass of the unbound population. In the upper panel, the position of the unbound particles is shown with blue points and the arrows represent the velocity vector associated with each particle. The medium panel shows a histogram of the position of the particles around the points $x=0$ (black points), $y=0$ (grey points), and $z=0$ (red points). The bottom panel is a histogram of the position angle relative to the $x$ axis.}
    \label{fig:Fig13}
\end{figure}

In summary, the gravitational interaction of a fast and massive particle with a compact particle cluster of low-mass particles can produce isotropic explosive outflows in a wide range of cluster masses, ejecting them in a direction related to the impact parameter of each one, with a tendency to follow the bullet direction of motion. As the cluster mass increases, the outflow spreads around the collision center but is still not as symmetric as the observed EOs. This could be related to the geometry of the cluster which could be distributed in a disk-like structure around another massive object, such as expected in another protostar. Therefore, it is necessary to explore different geometries, and mass-velocity distributions in future papers. {
A limitation to our model is that it is not directly explaining the formation of the gaseous filaments produced in the observed EOs. Then, we have analyzed the particle cluster evolution by only 5 yr, which is less than a hundredth the age of Orion~BN/KL.}
The mechanism proposed in this work can be compared to three mechanisms already proposed in the literature. The first one consists in a protostellar merger, that is, two protostars frontally collide, releasing their internal energy and accelerating their forming gas, but there is no detailed mechanism to eject material isotropically \citep{Bally2015}. The second one assumes that a young high mass and short lived star explodes as a supernova, releasing its forming gaseous planets, initially distributed in a low opening angle disk, and their wakes create the EO filaments \citep{Raga2021}. In this case, there is no evidence that a supernova remnant is present around the explosion center. The third proposal states that dynamic instabilities can be observed as a wind propagates from the ejection center, but again, there is no source associated around that point \citep{Dempsey20}.  

At the moment, our gravitational release model accounts for the more plausible mechanism to explain the initial dynamics of the EO, since it can explain the large number of fingers. Furthermore, using Equation \ref{eq:evaporation}, we can estimate the mass of the unbound population as $m_u=m_c \beta(m_c)$,  and taking the $8$M$_\odot$~Orion BN/KL mass as a lower limit for the ejected mass, we can estimate $m_u \ge 8$\msun. For $m_c<20$\msun, the function $m_u$ increases smoothly, getting larger than 8\msun ~at $m_c \sim9$\msun,  then 
$m_u$ reaches a local maximum at $m_c\sim 12$\msun, and finally $m_u$ decreases, getting smaller than 8\msun~ at $m_c\sim 15$\msun. In summary, the condition  $m_u \ge 8$\msun ~is met when $9{\rm M}_\odot\le m_c\le 15 {\rm M}_\odot$~which implies that the particle mass $m_i$ is in the sub-stellar mass scale. From Figure \ref{fig:F6}, the velocity in this mass interval is around $0.6v_{*0}$, which in our model corresponds to 60 km~s$^{-1}$ and the EO energy corresponds to $\sim 3 \times 10 ^{47}$erg, in agreement with the energy reported by \cite{Bally2024}. 

We have taken several reasonable assumptions and, therefore, our model has some caveats to be explored. First, we arbitrarily decided to use a 200-particle model to approximate the cluster to a continuum mass distribution and to resemble the large number of fingers shown by Orion BN/KL. Also, we assumed virial equilibrium and circular orbits as an initial condition for the initial velocity distribution, which is not the general case in star-forming regions. We ignored the gravitational binding energy of the parental cloud to obtain an approximation of the ejection velocity distribution. Still, a complete analysis of the dynamical evolution of the unbound cluster should consider the parental cloud potential and larger timescales for our simulations. Finally, we used, for simplicity, the same mass for each particle, when they might also be following a mass distribution. Each of these caveats has been discussed throughout the paper, but a more realistic model should consider them in more detail.

In any case, there is a long way to go to demonstrate that it is the mechanism capable to create them. Any of these previously mentioned hypotheses need to account for the ejection of several protostars and the formation of Hubble-like filaments that can be traced in H$_2$ and CO.

\begin{figure}
    \centering
    \includegraphics[width=\columnwidth]{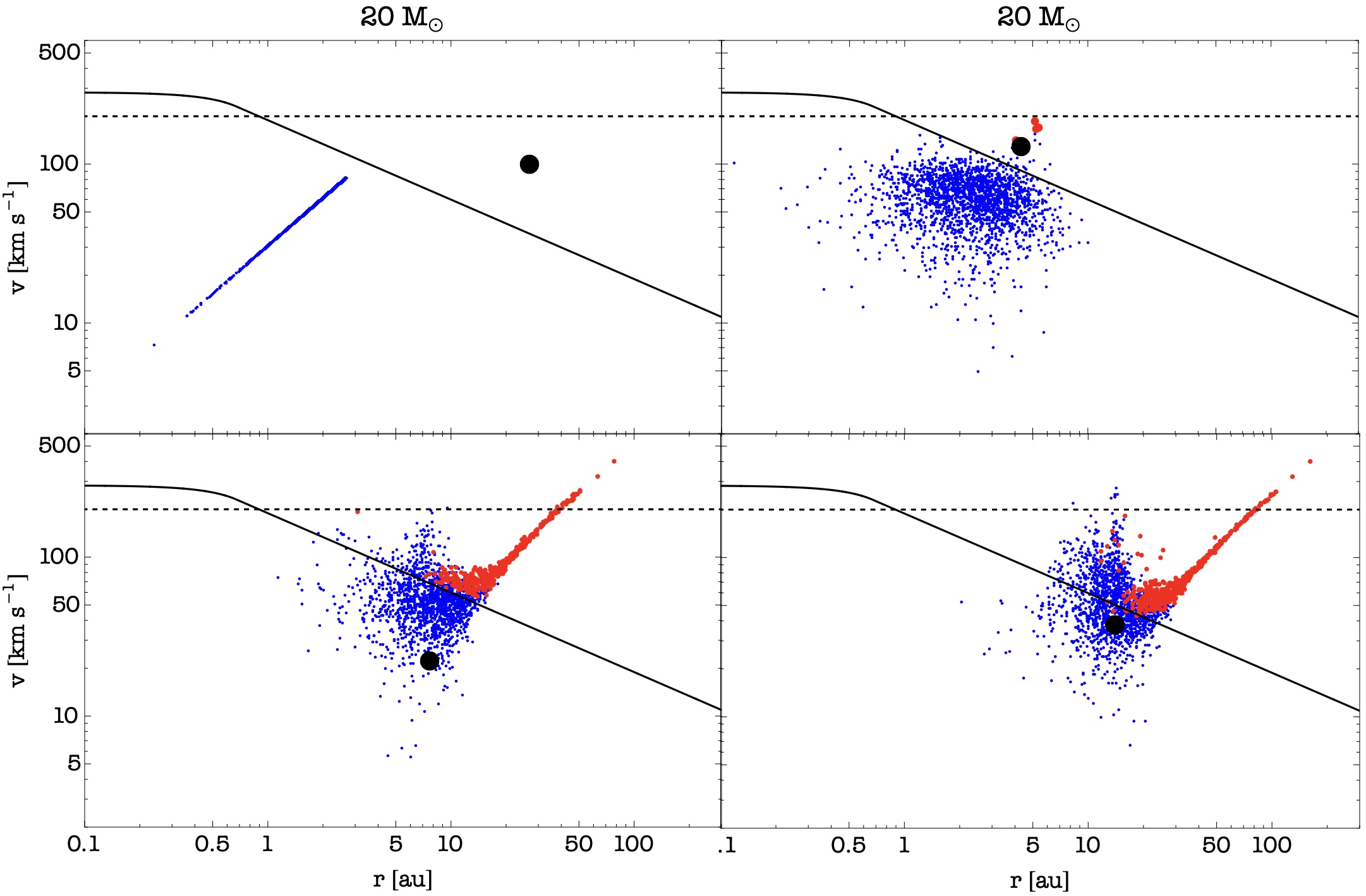}
    \caption{Phase diagram for the model with m$_c$=20 M$_\odot$ for 0, 0.5, 2, 3 yr  (top-left, top-right panel, bottom-left and bottom-right, respectively). The blue and red points represent the bound and unbound particles.  The continuous line represents the initial scape velocity as a function of $r$ , and the dashed line is the maximum velocity reached by the particles in the massless cluster model.}
    \label{fig:F15}
\end{figure}

%\subsection{Stellar dynamical explosion objects in the Galaxy}

%\textcolor{red}{Manuel y Estrella}
%MFL: Work in progress... I'll start working on this by mid-June. Right now it is just not possible for me. Sorry.
% \subsection{Other galactic dynamical explosion object}
% %The last of our models can be easily compared with the dynamic explosion presented in Orion BN/KL. As in figure 9 of paper I, we are interested in showing the effect of the dynamic interaction due to the passage of a massive star with high velocity. 
% %\newpage

\section{Conclusions}
\label{sec:sec6}
We have demonstrated the consequences of the collision between a star mass object moving at 100 km s$^{-1}$ directly towards a {  particle} clump cluster in gravitational equilibrium {  using $N$-body models of clusters between $10^{-3}$ and 50 \msun}, that could interact with the environment to create the observed filaments in EOs. We have probed whether the impact supplies enough energy and momentum through a gravitational interaction to accelerate a fraction of the cluster to velocities larger than the escape velocity of the initial configuration.  We summarize our conclusion on the following points:

\begin{enumerate}
    \item A model of interaction between a cluster of particles and an energetic bullet was proposed and reproduced in $N$--body numerical simulations, varying the mass of the perturbed cluster.
    \item  An evaporation factor related to the energy acquired by each particle in this interaction was defined.
    \item The evaporation factor depends on the mass of the cluster rather than its structure, that is, the evaporation parameter is constant even if it is measured in each concentric shell.
    \item A Hubble flow is achieved as a result of the interaction between the evaporated particles.
    \item The velocity of the particles is slightly more than twice the initial velocity of the bullet, which is the expected limit for the cluster of negligible mass.
    \item As the masses of the bullet and the cluster become more similar, the resulting flow produced by the evaporated particles becomes more isotropic from a static frame of reference. From the unbound frame of reference, the produced outflows are isotropic.
\end{enumerate}

  This is a first approximation to explain the mechanism that can produce EO, that has been associated with massive star-forming regions and more considerations about the cluster geometry, mass distribution and the interaction of the released particles with the environment need to be considered in the future implementing hydrodynamic simulations.

\section*{Acknowledgements}

We thank the anonymous referee for their useful comments that improved the clarity of this paper. We acknowledge support of the UNAM-PAPIIT grants IG101125, IN110722, IN103921, IN113119, IG100422, CONACYT grant 280775 and also the Miztli-UNAM supercomputer project LANCAD-UNAM-DGTIC-123 2022-1 and LANCAD-UNAM-DGTIC-128 2023-1.
E.G.C. and M.F.L. are truly grateful for all the support from UNAM’s Instituto de Radioastronomía y Astrofísica of Morelia. M.F.L. has received funding from the European Union’s Horizon 2020 Research and Innovation Programme under the Marie Sklodowska-Curie grant agreement No 734374 (LACE- GAL).

%%%%%%%%%%%%%%%%%%%%%%%%%%%%%%%%%%%%%%%%%%%%%%%%%%
\section*{Data Availability}
The data underlying this article will be shared on reasonable request to the corresponding author.

%%%%%%%%%%%%%%%%%%%%%%%%%%%%%%%%%%%%%%%%%%%%%%%%%%%%%%%%%%%%%%%%

%\appendix

%\section{Appendix information}

%\section{Author publication charges} \label{sec:pubcharge}

%\section{Rotating tables} \label{sec:rotate}

%%%%%%%%%%%%%%%%%%%%%%%%%%%%%%%%%%%%%%%%%%%%%%%%%%%%%%%%%%%%%%%%

\appendix

\section{Cluster velocity}
The constant $\alpha$ is a dimensionless parameter that can be determined if we consider the limit $\mu \to 0$ in Equation~\ref{eq:uc} ($u_c=\alpha$), and the case of a bullet colliding with a massless cluster. In general, the center of mass velocity of the dispersed cluster is

\begin{equation}
    \left< u_c \right>=\int_{u_{min}}^{u_{max}} u_x f(u)du,
    \label{eq:A1}
\end{equation}

 and, by \cite{Rivera2021}, the velocity projection of a particle moving in the direction $x$ is $u_x=u\cos\theta_\infty=u^2/2$ and the distribution function for such a cluster is given by

%\begin{equation}
%    f(u)du=\frac{12}{\xi'_c^3 u^3}(\xi'_c^2-{\xi'}^2)^{1/2}du
%\end{equation}

% which combined with 10 can be written as

 \begin{equation}
      f(u)du=\frac{12}{{\xi'_c}^3 u^4}\left[(1+\xi_c^{'2})u^2-4\right]^{1/2}du
      \label{eq:fu}
 \end{equation}

where $\xi_c^{'}$ is the size of the cluster in units of the gravitational radius $ Gm_*/v_{*0}^2$. Equation \ref{eq:fu}  can be combined with Equation \ref{eq:A1} to obtain 

\begin{equation}
     \left< u_c \right>=\frac{6}{{\xi'_c}^3}\int_{\frac{2}{(1+\xi_c^{'2})^{1/2}}}^{2}\frac{1}{u^2}\left[ \left( 1+ \xi_c^{'2}\right) u^2-4   \right]^{1/2}du
\end{equation}

Evaluating this integral, in the limiting case of a massless cluster, the velocity acquired by the center of mass of the cluster is $u_c=0.61$ and, therefore, $\alpha=0.61$.

\end{document}